\documentclass[
    ,final            
  ]
  {aipproc}

\layoutstyle{8x11single}

\begin{document}

\title{Electric dipole moments of nitric acid-water complexes measured by cluster beam deflection}


\author{Ramiro Moro}{
  address={Department of Physical Sciences, Cameron University,
Lawton, Oklahoma 73505, USA}
}

\author{Jonathon Heinrich}{
  address={Department of Physics and Astronomy, University of Southern California,
Los Angeles, California 90089-0484, USA} ,altaddress={Present
address: Department of Physics and Astronomy, University of Iowa,
Iowa City, Iowa 52240, USA} }

\author{Vitaly V. Kresin}{
  address={Department of Physics and Astronomy, University of Southern California,
Los Angeles, California 90089-0484, USA} }

\begin{abstract}
Water clusters embedding a nitric acid molecule
HNO$_3$(H$_2$O)$_{n=1-10}$ are investigated via electrostatic
deflection of a molecular beam. We observe large paraelectric
susceptibilities that greatly exceed the electronic polarizability,
revealing the contribution of permanent dipole moments.  The moments
derived from the data are also significantly higher than those of
pure water clusters.  An enhancement in the susceptibility for
$n$=5,6 and a rise in cluster abundances setting in at $n$=6 suggest
that dissociation of the solvated acid molecule into ions takes
place in this size range.
\end{abstract}

\maketitle


\section{Introduction}

One of the most basic phenomena in chemistry is the proton
delocalization that occurs when an acid is solvated in water. Strong
acids such as HCl are stated to dissociate completely in water, the
chlorine ion Cl$^-$ and the proton H$^+$ separating in the process.
Svante Arrhenius, who proposed this electrolytic theory of
dissociation, was recognized in 1903 with the Nobel Prize in
chemistry.  It is understood that the polar water molecules surround
the acid and weaken the electrostatic attraction between the
positive and negative ions to the point that dissociation becomes
energetically favorable. It is an interesting problem to study the
onset of this process in water clusters. For example, the minimum
number of water molecules required to promote dissolution of HBr was
determined experimentally to be five \cite{Hurley2002,
Robertson2002} and more recently molecules of HCl were observed to
dissociate in the presence of only four water molecules even in the
ultra-cold environment of a helium nanodroplet \cite{Gutberlet2009,
Zwier2009}.

The electric dipole moment of solvated charge-separated species is
expected to be higher than that of the intact molecule
\cite{Castleman1981}.  However, even large dipole moments will
manifest themselves only as thermally-averaged paraelectric
susceptibilities [see Eq. \eqref{Langevin} below] when observed at
high rotational and/or fluxional temperatures.  Thus, for example,
in electric-field deviation experiments molecular beams of polar
species commonly display weak deflection behavior mimicking that of
polarizable (i.e., induced-dipole) particles \cite{Broyer2002,
Handbook}. Consequently, instead of revealing its dipole moment
directly, the onset of dissolution will instead appear in a
deflection measurement as a change in the electric susceptibility
(effective polarizability). For example, a recent deflection study
of sodium-doped water clusters successfully detected the separation
of the 3\emph{s} electron and the Na$^+$ ion via this approach
\cite{Carrera2009}. An early attempt to observe the dissociation
transition of a hydrated HNO$_3$ molecule in a beam was carried out
in 1982 \cite{Kay1982}, but the experimental technique employed a
Stark focusing multipole, which is effective for detecting permanent
dipoles at low rotational temperatures in low-field-seeking states,
but not sensitive to deflections characteristic of induced
polarizabilities, which are both weaker and high-field-seeking.

Here we present the results of a beam deflection measurement of the
electrical susceptibilities of water clusters containing one nitric
acid molecule, and discuss the cluster electric dipole moments
derived from these results.

\section{Experiment}

The beam deflection apparatus was previously used to study the
electric deflection of water molecules and pure water clusters \cite
{Moro2006a, Moro2007}.  To prepare the mixed acid-water clusters, a
solution of nitric acid in water is heated inside a stainless steel
reservoir, the resulting vapor is mixed with helium that is supplied
to the reservoir from an external tank, and the mixture is allowed
to expand into a vacuum chamber through a 75 $\mu$m diameter nozzle.
The adiabatic cooling of the vapor-helium mixture results in the
coalescence of molecules into clusters. The produced distribution of
species includes pure clusters of water as well as mixed ones
containing one, two or more acid molecules. The beam passes through
a skimmer into a second vacuum chamber, where it is collimated and
then travels through the region of an inhomogeneous "two-wire"
\cite{Tikhonov2002} electric field created by a 15 cm long pair of
metal electrodes.  By applying voltages of up to 27.5 kV between the
electrode plates, electric fields and field gradients of up to 80
kV/cm and 380 V/cm$^2$, respectively, are created. The field induces
an alignment of the clusters' dipole moments, and the gradient gives
rise to a net deflecting force on the beam.

The deflected clusters then pass through a rotating chopper and
travel to a quadrupole mass analyzer where they are ionized by
bombardment with 70 eV electrons, mass selected, and detected by an
analog multiplier.  The detector output is fed into a lock-in
amplifier together with the chopper synchronization pulses.  This
eliminates noise from the background gases in the chamber and allows
determination of the beam velocity from the phase difference between
the chopping and detection signals (average $v$=1130 m/s in these
experiments).

To determine the beam profiles, a 0.25 mm wide slit is positioned in
front of the detector entrance, a distance $L$=71 cm past the middle
of the deflection field plates. The slit is scanned across the beam
by a stepper motor in steps of 0.25 mm, measuring the intensity at
each position with and without voltage applied to the deflection
plates.

One of the challenges in a molecular beam experiment like this is to
determine the composition of parent clusters from the mass spectra
of the ions.  Sometimes there is no straightforward way to do so:
e.g., in the case of HCl solvated in water clusters the chlorine
atom always departs upon ionization \cite{Kay1982, Moro2006b}, so
the mass spectrum is indistinguishable from that originating from
pure water. Fortunately, in the case of mixed HNO$_3$(H$_2$O)$_n$
clusters it has been found \cite{Kay1982} that the acid is retained
upon electron impact ionization, the main fragmentation channel
being the loss of one hydroxyl group (just as in the case of pure
water clusters). This permits easy identification of the neutral
parent.

\section{Results and discussion}

\begin{figure}
  \includegraphics[height=.4\textheight]{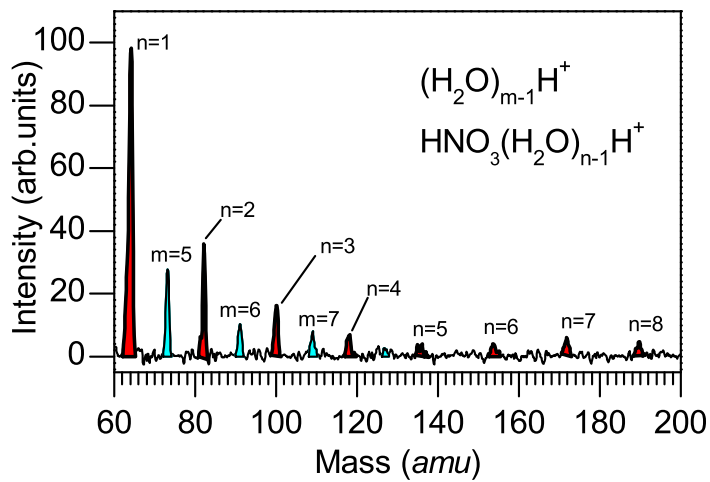}
  \caption{Mass spectrum produced by electron ionization of a
molecular beam formed by nozzle expansion of a solution of nitric
acid in water (concentration 11.2 M) kept in a source reservoir at
70$^o$C and accompanied by helium carrier gas.  Indices $m$ and $n$
denote the sizes of the original neutral parent clusters in the
beam.  Note the non-monotonic trend of the mixed acid-water cluster
intensities.} \label{Fig1}
\end{figure}

The first observation to consider is the distribution of cluster
intensities. Fig. \ref{Fig1} shows a typical mass spectrum obtained
in the present experiments. We observe (H$_2$O)$_{m-1}$H$^+$ ions
deriving from (H$_2$O)$_m$ as well as HNO$_3$(H$_2$O)$_{n-1}$H$^+$
peaks corresponding to HNO$_3$(H$_2$O)$_n$ parents that lost an OH
group in the ionization process.


The HNO$_3$(H$_2$O)$_n$ abundance spectrum reveals a minimum at
$n$=5. This feature has been observed previously \cite{Kay1982} and
interpreted as a possible signal of proton delocalization. The
rationale is that a dissociated acid molecule can enhance cluster
stability, hence yielding an upturn in intensity for $n\geq6$.

\begin{figure}
  \includegraphics[height=.4\textheight]{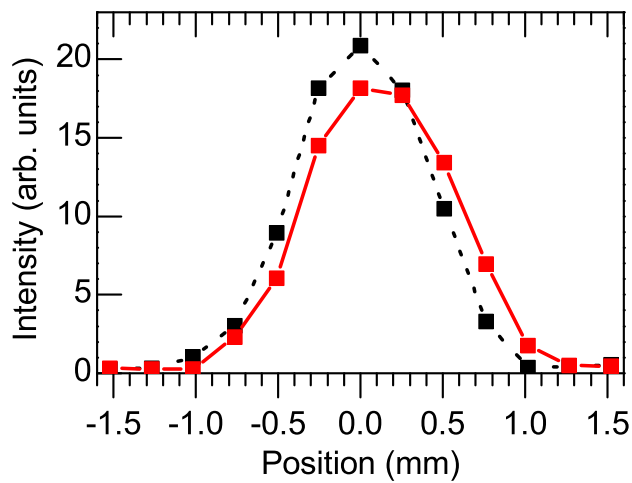}
  \caption{A typical cluster beam deflection profile.  The
illustrated data corresponds to HNO$_3$(H$_2$O).  Dashed line: beam
profile without electric field, solid line: with an applied electric
field of 80 kV/cm and a gradient of 380 kV/cm$^2$.  The field
produces a deflection towards the right (region of stronger electric
field), indicating a susceptibility equivalent to an effective
polarizability.} \label{Fig2}
\end{figure}

We now come to the results of the deflection measurement. As an
example of the experimental data, Fig. \ref{Fig2} shows the
deflection behavior of an agglomerate of one acid molecule and one
water molecule. The dashed line corresponds to the beam profile when
no electric field is applied. The width of the profile reflects the
actual size of the beam due to the finite dimensions of the skimmer
and collimator. Once the electric field is switched on, the beam is
deflected towards the right, i.e., towards the region of a stronger
field.  Such uniform deflection without broadening is characteristic
of "floppy" species (that undergo internal rotational/vibrational
fluctuations, interconverting between different conformations and
correspondingly different orientations of their constituent
dipoles), as well as of asymmetric rotors characterized by congested
rotational Stark spectra \cite{Broyer2002, Handbook, Xu2008}.

From the magnitude of the deflection, $d$, one can derive the
effective polarizability values for the clusters, $\alpha^{eff}$,
according to the standard equation

\begin{equation}
d = \alpha ^{eff} C\frac{{V^2 }}{{mv^2 }}, \label{deflection}
\end{equation}
where $m$ is the cluster mass, $v$ its velocity, $V$ is the applied
voltage and $C$ is a calibrated geometrical coefficient
\cite{Moro2006a}.

\begin{figure}
  \includegraphics[height=.4\textheight]{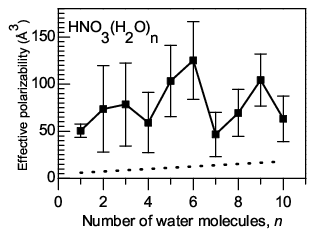}
  \caption{Effective electric polarizabilities of water clusters
containing one nitric acid molecule (squares).  Sum of the
electronic contributions to the polarizability from the water
clusters and the acid molecule is shown for comparison as a dashed
line.} \label{Fig3}
\end{figure}

The results for the HNO$_3$(H$_2$O)$_{n=1-10}$ cluster series are
shown in Fig. \ref{Fig3}.  The sum of the electronic
polarizabilities of HNO$_3$ and (H$_2$O)$_n$ are also displayed for
comparison. For water $\alpha^{electronic}$ was taken from the
calculations in Ref. \cite{Yang2005}, and for the nitric acid
molecule the estimate given in Ref. \cite{Huey1996} was used.

Fig. \ref{Fig3} makes it clear that the effective polarizabilities
greatly exceed the molecular electronic contribution, revealing the
presence of permanent dipole moments.  A transitional pattern can be
observed in the $n$=4-7 range.  The absolute ensemble-averaged
dipole moments of the clusters, $\bar p$, can be extracted from the
data with the use of the Langevin linear response equation,
analogous to that used for pure water clusters and a number of other
species \cite{Broyer2002, Handbook, Moro2006a}:

 \begin{equation}
\alpha ^{eff}  = \alpha ^{_{electronic} }  + \frac{{\bar p^2
}}{{3k_B T}}.
 \label{Langevin}
\end{equation}

We use a cluster temperature of $T$=170K, which was determined for
the conditions of the present experiment from the deflection profile
of the water molecule \cite{Moro2007}, and also found to be
consistent with the susceptibility of the pure water pentamer
\cite{Moro2006a}.

\begin{figure}
  \includegraphics[height=.4\textheight]{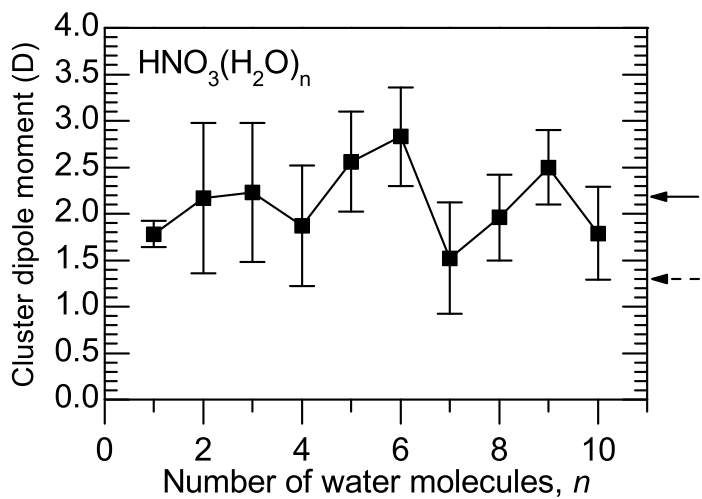}
  \caption{Electric dipole moments of mixed clusters deduced using
the data in Fig. \ref{Fig3} and Eq. (\ref{Langevin}).  The dipole
moment of the gas-phase nitric acid molecule (2.17 D) is shown as
the solid arrow \cite{CRC} and the dipole moment of pure water
clusters in the size range $n$=3-8 is shown as the dashed arrow
\cite{Moro2006a}.} \label{Fig4}
\end{figure}

The derived magnitudes of the dipole moments are shown in Fig.
\ref{Fig4}. It is important to recognize that the values for
HNO$_3$(H$_2$O)$_n$ significantly exceed those of pure (H$_2$O)$_n$
clusters under similar conditions ($\approx$1.3 Debye for $n$=3-8
and $\approx$1.6 D for $n$=9-10 \cite{Moro2006a}).

A comparison with computations available in the literature
\cite{Escribano2003, Scott2004} shows that the experimental results
are considerably smaller than the predicted dipole moments, see
Table \ref{Table1}.  A likely explanation is that whereas
theoretical calculations predominantly concentrate on the low-energy
isomers of pure and doped water clusters, actual physical systems
exist at temperatures high enough that their strongly fluxional
nature explores a large number of configurations, producing a
reduced average magnitude of the dipole moment.  Challenging but
important steps to investigate this thermal isomer dynamics would
involve theoretical calculations for more realistic conditions, as
well as experimental efforts to investigate such free polar species
at low temperatures.

\begin{table}[tbh]
\begin{tabular}{lccc}
\hline
  & \tablehead{1}{c}{c}{Theory (Ref. \cite{Escribano2003})}
  & \tablehead{1}{c}{c}{Theory (Ref. \cite{Scott2004})\tablenote{Range of values corresponding to several different low-energy isomers}}
  & \tablehead{1}{c}{c}{Present results}
  \\
\hline
HNO$_3$(H$_2$O)     & 4.2 D & 3.8-4.4 D & 1.8$\pm$0.2 D\\
HNO$_3$(H$_2$O)$_2$ & 3.8 D & 3.9-4.4 D & 2.2$\pm$0.8 D\\
HNO$_3$(H$_2$O)$_3$ & 3.6 D & 2.6-3.8 D & 2.2$\pm$0.8 D\\
HNO$_3$(H$_2$O)$_4$ &       & 0.8-1.8 D & 1.9$\pm$0.7 D\\
\hline
\end{tabular}
\caption{Electric dipole moments of small mixed clusters}
\label{Table1}
\end{table}

What about signatures of acid dissociation in HNO$_3$(H$_2$O)$_n$?

As mentioned above, it has been suggested that the uptick in mixed
cluster abundances at $n$=6 water molecules may correspond to such a
transition.

On the other hand, Scott et al. \cite{Scott2004} suggest that the
minimum number of water molecules that admit dissociation is $n$=4,
with H$_3$O$^+$ and NO$_3^-$ ions stabilized by three water
molecules that separate them.  This cannot be excluded because we do
observe a decrease in the dipole moment at $n$=4, but the decrease
is small, and the magnitude of the dipole suggests that in actuality
a more varied set of configurations may be involved, as commented in
the previous paragraph.

The aforementioned alternative expectation that the electric dipole
moment of the system should increase when the species becomes
charge-separated \cite{Castleman1981} points instead at a critical
size of $n$=5, when there is a noticeable increase in the
susceptibility and the corresponding dipole moment.  One could argue
that adding further water molecules enables them to start
surrounding the ions, thereby screening them, reducing the effective
dipole moment, and causing the decrease in $\bar p$ observed for
larger species.

Although the present measurement does not point to an unequivocally
precise number of water molecules corresponding to the departure of
the proton from solvated HNO$_3$, one also has to keep in mind that
such a solvation step does not have to occur as an abrupt step.  It
can certainly represent a more gradual structural transition.  The
abundance and deflection data presented here provide evidence that
the acid dissociation takes place in the range between $n$=5-6 in
free water clusters of internal temperature of $T\approx$170 K.  It
would be very informative to consider, experimentally and
theoretically, whether and how this process varies with $T$.

\begin{theacknowledgments}
 We would like to thank Dr. D. McGuire for a useful discussion.
This work was supported by the U.S. National Science Foundation
(PHY-0652534) and by the USC Undergraduate Research Associates
Program.
\end{theacknowledgments}

\end{document}